\documentclass[lettersize,journal]{IEEEtran}

\usepackage{amsmath,epsfig,amssymb,verbatim,amsopn,cite,multirow}
\usepackage{balance}
\usepackage{multirow}
\usepackage[usenames,dvipsnames]{color}
\usepackage[all]{xy}  
\usepackage{url}
\usepackage{amsfonts}
\usepackage{amssymb}
\usepackage{epsfig}
\usepackage{epstopdf}
\usepackage{bm}
\usepackage{balance}
\usepackage{graphicx}
\usepackage{subcaption}

\usepackage{cancel}
\usepackage{algorithm}
\usepackage{dblfloatfix}
\usepackage{algpseudocode}
\usepackage{algorithm}
\usepackage{algpseudocode}

\usepackage{amsthm}

\newtheorem{remark}{Remark}

\usepackage[margin=14.5mm,top=18.1mm,bottom=25.7mm]{geometry}

\usepackage[nodisplayskipstretch]{setspace}
\newcounter{mytempeqcounter}
\usepackage{pifont}

\newcommand{\qb}{{\bf b}}

\newcommand{\qf}{{\bf f}}
\newcommand{\qg}{{\bf g}}
\newcommand{\qh}{{\bf h}}

\newcommand{\qn}{{\bf n}}

\newcommand{\qr}{{\bf r}}

\newcommand{\qw}{{\bf w}}
\newcommand{\qx}{{\bf x}}

\newcommand{\qA}{{\bf A}}

\newcommand{\qC}{{\bf C}}

\newcommand{\qG}{{\bf G}}

\newcommand{\qJ}{{\bf J}}

\newcommand{\qL}{{\bf L}}

\newcommand{\qW}{{\bf W}}
\newcommand{\qX}{{\bf X}}

\makeatletter
\renewcommand{\@makefnmark}{%
  \hbox{\@textsuperscript{\normalfont\@thefnmark}}}
\makeatother

\begin{document}

\title{ Distributed Pinching-Antenna Architecture for Communication and Target Positioning}

\author{Haoran Ni,  Jiajun He, Mohammadali Mohammadi, Hien Quoc Ngo, and Michail Matthaiou\\
Centre for Wireless Innovation (CWI), Queen's University Belfast, U.K.\\
Email: \{hni02, j.he, m.mohammadi, hien.ngo, m.matthaiou\}@qub.ac.uk 
}


\maketitle
\thispagestyle{empty}
\begin{abstract}
We propose a pinching-antenna system (PASS)-assisted integrated sensing and communication (ISAC) framework for joint downlink communication and target positioning. The dielectric waveguides enable flexible signal radiation through activated pinching antennas (PAs), while the reflected echo signals are collected by a sensing receiver. A unified waveform is adopted to support both data transmission and sensing-echo processing. In particular, a two-stage procedure is developed for target localization using time-sum-of-arrival (TSoA) measurements, where the signal propagation distances are first estimated through de-chirp processing and beat frequency extraction. Then, the target position is obtained based on TSoA measurements using the weighted linear least squares(WLLS) technique. To characterize the fundamental limits of the PASS-assisted ISAC system, a Cramér-Rao lower bound (CRB) is derived, which captures the impact of PA activation. Furthermore, the PA activation problem is formulated to jointly  optimize the communication spectral efficiency (SE) and sensing performance. To address this, a low-complexity heuristic greedy selection algorithm is proposed to determine the activated PA pattern. Numerical results demonstrate that the proposed scheme outperforms random PA selection, achieving gains of up to $81\%$ improvement in communication SE.
\end{abstract}


\IEEEpeerreviewmaketitle

\vspace{-1em}
 \section{Introduction}
{\renewcommand{\thefootnote}{}%
\footnotetext{This work was supported by the U.K. Engineering and Physical Sciences Research Council (EPSRC) grant (EP/X04047X/2) for TITAN Telecoms Hub, the European Research Council (ERC) under the European Union’s Horizon 2020 research and innovation programme (grant agreement No. 101001331), and  the MULTIPLY-6G project that has received funding from the Smart Networks and Services Joint Undertaking (SNS JU) under the European Union’s Horizon Europe research and innovation programme under Grant Agreement No 101293106. The work of J. He was supported by the HORIZON-MSCA through the project ECO-ISAC under Grant 101274088. The work of H. Q. Ngo and M. Matthaiou was supported by a research grant from the Department for the Economy Northern Ireland under the US-Ireland R\&D Partnership Programme. }}

\IEEEPARstart{I}{SAC} has emerged as a promising paradigm for next-generation wireless networks, as it enables communication and sensing functionalities to share spectrum, hardware, and signal resources. Meanwhile,  flexible-antenna technologies, such as fluid and movable antennas, have been developed to improve spatial adaptability, but they still face inherent limitations in mitigating severe large-scale fading\cite{zhou2024fluid}. 

To overcome these limitations, PASS architectures have  attracted increasing attention for improving the performance of ISAC systems, thanks to their capability of flexibly creating line-of-sight (LoS) links for both reliable communication and high-precision sensing\cite{11391537}. By activating different PAs along the dielectric waveguides, a PASS can reshape the effective propagation paths and provide additional spatial degrees of freedom for joint communication and sensing.

Recent studies have explored PASS variants for different ISAC configurations and performance objectives. In \cite{11111701}, a wireless sensing architecture based on PASS was proposed, where leaky coaxial cables were used for collecting echo signals, and the waveform and PA positions were jointly optimized to improve multi-target sensing accuracy. In particular, recent work investigated PA-assisted positioning from a CRB perspective, showing that the reconfigurable and flexible array configuration of PAs can improve localization accuracy and support flexible user-centric positioning\cite{ding2025pinching}. A received signal strength-based localization framework for PASS was developed in \cite{11372748}, where stochastic geometry was used to characterize target localizability and CRB performance. By exploiting spatial-phase diversity through non-uniform PA placement and in-waveguide propagation, PASS-ISAC can significantly tighten the sensing bounds compared with a uniform linear array \cite{bozanis2025cramer}. The study in \cite{xue2026pinching} investigated a PASS-ISAC framework with asymmetric sensing propagation, derived the sensing degrees of freedom and angle estimation CRB, and optimized the PA locations and transmit power to balance sensing accuracy and power consumption.

Although existing studies have advanced the understanding of PASS-assisted ISAC systems, several challenges remain underexplored for joint communication and localization. First, due to the limited number of radio-frequency (RF) chains in practical PASS architectures, PAs connected to the same dielectric waveguide typically transmit the same signal, necessitating a simple unified waveform that simultaneously supports communication and sensing \cite{11185315}. Second, PA activation affects not only sensing-link quality but also the localization degrees of freedom determined by the sensing geometry. Thus, its impact on localization accuracy should be explicitly characterized from a geometry-based perspective. Third, the role of PA activation in balancing communication performance and localization accuracy remains insufficiently explored.

Motivated by these challenges, this paper proposes a PASS-assisted ISAC localization framework with PA activation design. The
main contributions of this paper are  as follows:
\vspace{-0.2em}
\begin{itemize}
    \item A unified waveform supports both communication and sensing under a RF chain-limited PASS architecture. This design reduces waveform complexity and avoids dedicated sensing signals. The transmitted signal can support data transmission and echo-based range estimation, which is suitable for PASS architectures with shared RF chains.
    \item We reveal that the PA positions affect not only the effective communication and sensing channels, but also the sensing geometry for target localization. Based on this observation, a geometry-based CRB analysis is developed to characterize the achievable localization accuracy under different PA activation patterns.
    \item We formulate a PA activation optimization problem that balances between communication SE and sensing performance while ensuring localization feasibility. To solve this discrete and nonconvex problem efficiently, a low-complexity heuristic greedy selection algorithm is proposed to determine the activated PA pattern.
\end{itemize}

\noindent\textit{Notations:} Scalars, vectors, and matrices are denoted by lowercase, bold lowercase, and bold uppercase letters. The sets of real and complex numbers are $\mathbb{R}$ and $\mathbb{C}$. For a complex scalar $x$, $x^{*}$ and $|x|$ denote its conjugate and magnitude. For vectors or matrices, $(\cdot)^T$ and $(\cdot)^H$ indicate the transpose and conjugate transpose. The $n$-th entry of a vector $\mathbf{x}$ is $[\mathbf{x}]_n$, and $|\cdot|$ denotes the Euclidean norm. The operators $\operatorname{diag}(\cdot)$ and $\operatorname{blkdiag}(\cdot)$ form diagonal and block diagonal matrices, while $\operatorname{Tr}(\cdot)$ and $\operatorname{rank}(\cdot)$ denote the trace and rank. Finally, $\mathbb{E}\{\cdot\}$ represents expectation.

\vspace{-0.3em}
\section{System model}~\label{sec:model}
We consider a  downlink PASS system. The base station (BS) is connected to $M$ transmit dielectric waveguides of length $L$, each  containing  $N$  PAs. The PASS serves a single-antenna communication user (CU) and senses a point-like target. Both the CU and the target are located in a region of size $D=D_x\times D_y$ $\text{m}^2$.  The sensing receiver is a single-antenna access point (AP) that observes the reflected sensing signals. 
The dielectric waveguides are assumed to be parallel to the $x$-axis with respect to the $x-o-y$ plane at a height $d$. The $m$-th waveguide starts at the feed point $\bm{\psi}_{m,0}=[0,y_{m,0},d]^T$ and ends at  $[L,y_{m,0},d]^T$. In particular, the position of the $n$-th PA on the $m$-th waveguide can be denoted by $\bm{\psi}_{m,n}=[x^{\text{p}}_{m,n},y^{\text{p}}_{m,n},d]^T$. The $x$-axis coordinates of all PAs can be stacked into a matrix $\mathbf{X}=[\qx_1,\qx_2,\dots,\qx_M]\in\mathbb{R}^{N \times M}$, where $\qx_m=[x^{\text{p}}_{m,1},x^{\text{p}}_{m,2},\dots,x^{\text{p}}_{m,N}]^T$ represents the $x$-axis coordinate vector of PAs on the $m$-th waveguide.
The positions of the CU and the sensing target are denoted $\boldsymbol{\psi}^{\zeta}=[x^{\zeta},y^{\zeta},z^{\zeta}]^T$ with $\zeta\in\{c,s\}$. Similarly, the AP position is $\bm{\psi}^{\text{AP}}=[x^{\text{AP}},y^{\text{AP}},z^{\text{AP}}]^T$.


\vspace{-0.8em}

\subsection{Channel Model}
We consider standard indoor LoS-only scenarios in which the CU and target are in LoS with the PAs. 
The channel vector from the $m$-th waveguide to the sensing AP or CU is
\vspace{-0.2em}
\begin{equation}
\mathbf{h}_{\zeta,m}(\qx_m)
\!= \!
\left[
\frac{\sqrt{\eta} e^{-j \frac{2\pi}{\lambda} D_{m,1}^{\zeta}}}{D_{m,1}^{\zeta}},
\;\cdots,\;
\frac{\sqrt{\eta} e^{-j \frac{2\pi}{\lambda} D_{m,N}^{\zeta}}}{D_{m,N}^{\zeta}}
\right]^H,
\end{equation}
where $\lambda = \frac{c}{f_c}$ denotes the wavelength;
$\eta = \frac{c^2}{16\pi^2 f_c^2}$ with $c$ and $f_c$ being the speed of light and carrier frequency, respectively; $D_{m,n}^{\zeta}$ denotes the propagation distance associated with the $n$-th PA on the $m$-th waveguide. For the communication link, $D_{m,n}^{c}=\Vert\boldsymbol{\psi}^{c}-\boldsymbol{\psi}_{m,n}\Vert$, while for the sensing link, $D_{m,n}^{s}=\Vert\boldsymbol{\psi}^{s}-\boldsymbol{\psi}_{m,n}\Vert+\Vert\boldsymbol{\psi}^{s}-\boldsymbol{\psi}^{\mathrm{AP}}\Vert$.  Moreover, the overall  channel vector is given by
\vspace{-0.2em}
\begin{equation}
  \qh_{\zeta}(\qX)\!=\!\big[\mathbf{h}^H_{\zeta,1}(\qx_1),\dots,\mathbf{h}^H_{\zeta,M}(\qx_M) \big]^H\!\in\! \mathbb{C}^{MN\times1}.
\end{equation}

\begin{figure}[t]
    \centering
    \includegraphics[width=0.45\textwidth]{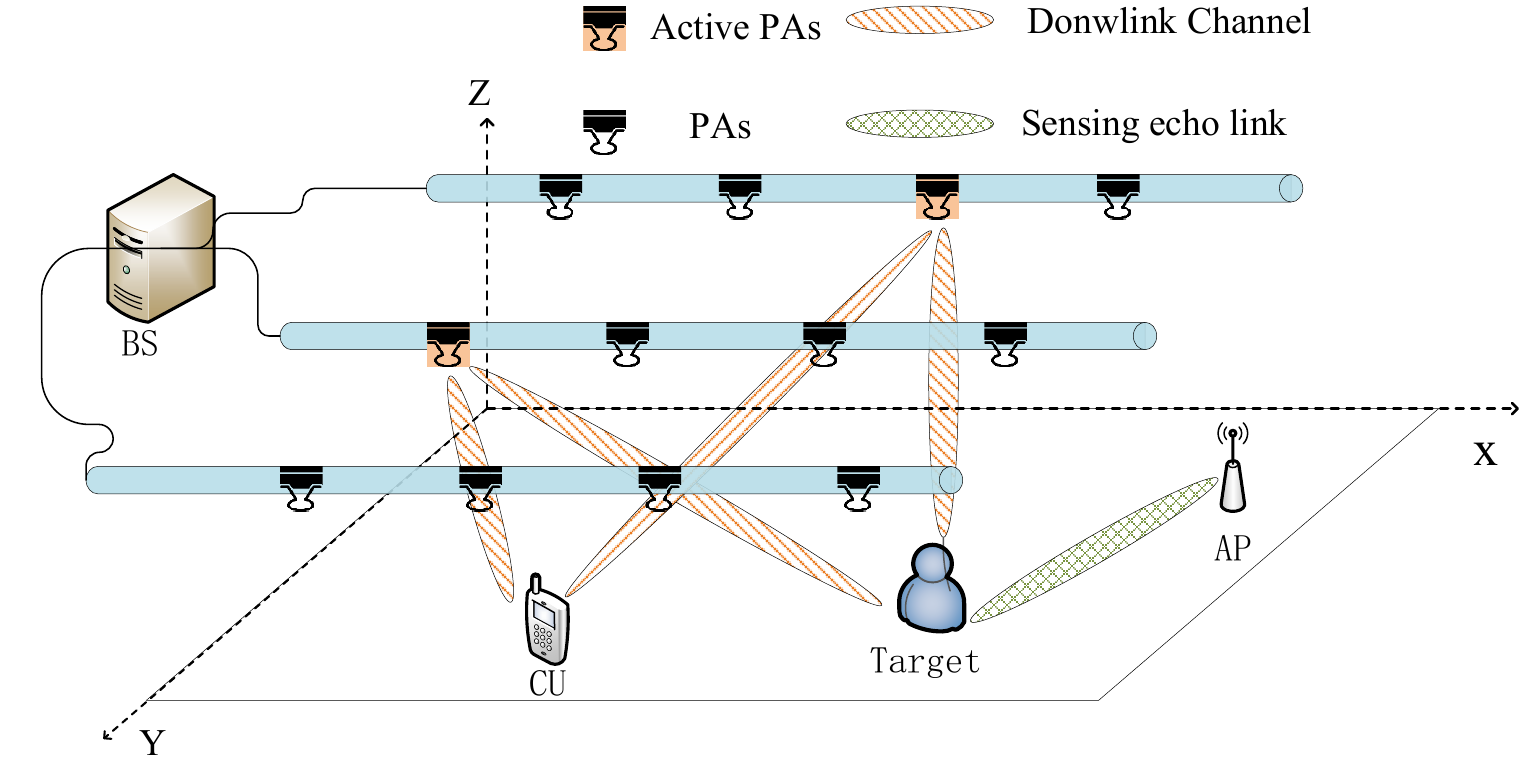}
    \caption{System model of the PASS-assisted ISAC.}
    \label{fig:placeholder}
    \vspace{0.2em}
\end{figure}

We define $\lambda_g=\frac{\lambda}{n_e}$ as the wavelength of the carrier in the waveguide, where $n_e$ is the effective refractive index\cite{pozar2021microwave}. Following \cite{11391537}, the in-waveguide attenuation can be neglected under mild conditions, as it induces a negligible loss in the average rate.
The in-waveguide channel vector  $ \qg_m(\qx_m)\in\mathbb{C}^{N\times1}$ for the $m$-th waveguide  can be expressed as
\begin{equation}
    \qg_m(\qx_m)\!=\!\left[ e^{\!-j\frac{2\pi}{\lambda_g}\Vert\bm{\psi}_{m,1}-\bm{\psi}_{m,0}\Vert},\dots,e^{\!-j\frac{2\pi}{\lambda_g}\Vert\bm{\psi}_{m,N}-\bm{\psi}_{m,0}\Vert}\right]^{\!T}\!.
\end{equation}
Therefore, the overall in-waveguide channel matrix can be represented by a block-diagonal matrix $\qG\in\mathbb{C}^{MN\times M}$\cite{11371592}, which is given by
 \vspace{-0.2em}
\begin{align}
\mathbf{G}(\mathbf{X})
&= \operatorname{blkdiag}\!\left(
\mathbf{g}_1(\mathbf{x}_1),
\mathbf{g}_2(\mathbf{x}_2),
\ldots,
\mathbf{g}_M(\mathbf{x}_M)
\right)\notag\\
&=
\begin{bmatrix}
\mathbf{g}_1(\mathbf{x}_1) & 0 & \cdots & 0 \\
0 & \mathbf{g}_2(\mathbf{x}_2) & \cdots & 0 \\
\vdots & \vdots & \ddots & \vdots \\
0 & 0 & \cdots & \mathbf{g}_M(\mathbf{x}_M)
\end{bmatrix}.
\end{align}
\vspace{-1.1em}
\subsection{PA Configuration Strategies}
We consider a binary activation strategy for the candidate PAs deployed on the waveguides. The PA activation vector is
\vspace{-0.2em}
\begin{equation}
\boldsymbol{\alpha} = [\alpha_1,\alpha_2,\ldots,\alpha_{m,n},\ldots,\alpha_{MN}]^T \in \{0,1\}^{MN\times 1},
\end{equation}
 where \(\alpha_{m,n} = 1\) indicates that the \(n\)-th PA on the \(m\)-th waveguides is active, and \(\alpha_{m,n} = 0\) otherwise. It is assumed that a total of \(K\) PAs are activated. Accordingly, the corresponding PA activation matrix is defined as $\mathbf{A} = \operatorname{diag}(\boldsymbol{\alpha}) \in \mathbb{R}^{MN \times MN}$.
With this activation strategy, only \(K\leq MN\) active PAs participate in the transmission and sensing processes.

 \vspace{-0.8em}
\subsection{Signal Model}
Linear frequency modulation (LFM) is adopted since it enables efficient sensing delay estimation by mapping delay to the beat frequency. In addition, the LFM signal provides a flexible carrier for communication symbol embedding\cite{lyu2023radar}. The adopted LFM signal can be expressed as
\vspace{-0.2em}
\begin{equation}\label{pure_LFM}
    s_{\text{L}}(t)=\text{exp}\big[j\pi(2f_0t+ut^2) \big],
\end{equation}
where $f_0$ is the initial frequency and $u$ is the chirp rate.
The LFM signal sweeps linearly over bandwidth $B$ within the temporal
duration $T$ with a rate of $u = B/T$. 
The LFM waveform serves  as the communication carrier, and the data sequence is mapped to the quadrature amplitude modulation (QAM) symbols in each time slot $T_s$\cite{10506595}. Thus, the integrated waveform is written as
\begin{equation}
    s(t)=a(t)\text{exp}[j\pi(2f_0t+ut^2)+j\phi(t)],
\end{equation}
where $a(t)$ and $\phi(t)$ are the amplitude and phase of the QAM symbols, respectively.
For 4-QAM, the constellation has a constant modulus and thus yields a constant-envelope waveform with $a(t)= 1$, which can also enhance the sensing performance\cite{10147248}. In contrast, higher-order QAM (e.g., 16-QAM) is multi-modulus and yields a non-constant envelope, which may improve the communication performance but can degrade the sensing accuracy\cite{10506595}. We focus on the constant-envelope case and set $a(t)=1$ over $t\in[0,T]$.

\vspace{-0.5em}
\section{Communication Receiver Processing and  Demodulation}

\subsection{CU Received Signal Model}
In the downlink, the CU receives a superposition of delayed integrated  waveforms radiated by the PAs. Let $\mathbf{w} = [w_1, \ldots, w_M]^T \in \mathbb{C}^{M \times 1}$ denote the transmit beamforming vector for the CU, designed using maximum ratio transmission (MRT). The received signal at CU at time  $t$ is given by
\begin{equation}\label{eq:CU_recivced}
y_c(t)
\!=\!
\sqrt{P_t}\!
\sum_{m=1}^{M}\!
\sum_{n=1}^{N}\!
\alpha_{m,n}
[\qh_{c,m}^{*}]_n
[\qg_{m}]_n
w_m
s\left(t\!-\!\tau_{m,n}^{c}\right) \!
+  n_c(t),
\end{equation}
where $P_t$ is the transmit power fed into the waveguides, $\tau_{m,n}^{\mathrm{c}}$ represents the propagation  delay determined by the distance between the $n$-th PA on the $m$-th waveguide and the CU, and  $n_c(t)$ denotes the additive white Gaussian noise (AWGN) at time slot $t$, whose power is $\sigma_c^2$. We point out that communication and sensing are supported by the same integrated waveform and no additional sensing interference is imposed on the CU. Accordingly, the SNR for the CU is $    \text{SNR}_c=\frac{P_t\vert \qh_c^H\qA\qG\qw \vert^2}{\sigma_c^2}.$
For PA activation design, we adopt the Gaussian-input SE benchmark $\text{SE}=\log_2(1+\text{SNR}_c)$.
\enlargethispage{-0.03in}
\vspace{-1em}
\subsection{Signal De-chirp and Restoration Processing}
 Since the carrier parameters of the LFM waveform are known at the CU, the local reference signal in \eqref{pure_LFM}  can be generated for de-chirp processing with \eqref{eq:CU_recivced} by
\vspace{-0.2em}
\begin{align}\label{eq:zc_def}
z_c(t)&=y_c(t)s_{\text{L}}^{\ast}(t)\notag\\
&\!=\!\sum\nolimits_{m=1}^{M}\!\sum\nolimits_{n=1}^{N}\!\!\xi_{m,n}^{c}e^{j\phi(t-\tau_{m,n}^c)}e^{-j2\pi u\tau_{m,n}^c t}
\!+\!\tilde n_c(t),
\end{align}
where $\xi^c_{m,n}=\sqrt{P_t}\alpha_{m,n}
[\qh_{c,m}]_n^{*}
[\qg_{m}]_n
w_m$ is the  complex gain and phase terms of the cascaded in-waveguide and free-space propagation, while $\tilde n_c(t)=n_c(t)s_{\text{L}}^{\ast}(t)$.  The data phase remains unchanged during each symbol interval of duration $T_s$. Let $I=\frac{T}{T_s}$ denote the number of symbol intervals within one LFM waveform duration, and  $i\in\{1,\ldots,I\}$ denote the symbol index. Since the data phase is constant within each symbol interval and $\tau_{m,n}^{c}\ll T_s$, the delayed phase term $\phi(t-\tau_{m,n}^{c})$ can be approximated by the same symbol phase $\phi_i^c$ over the $i$-th interval. Since the propagation delay satisfies $\tau_{{m,n}}^{c}\ll T_s$, the delayed phase term is still the same symbol. Therefore, for the $i$-th symbol interval, it can be approximated as a constant phase term, denoted by $\phi_i^c$. We rewrite  \eqref{eq:zc_def} as
\begin{equation}
    z_c(t)=\sum\nolimits_{m=1}^{M}\sum\nolimits_{n=1}^{N}\xi^{c}_{m,n}e^{j\phi_{i}^c}e^{-j2\pi u\tau_{m,n}^ct}
+\tilde n_c(t),
\end{equation}
which is a standard multi-path continuous-time signal model.
With de-chirp processing, the delay-dependent phase can be estimated and compensated within each symbol interval, after which standard QAM demodulation can be performed based on the restored symbol phase\cite{sui2025multifunctional}.

\vspace{-0.1em}
\section{Sensing Signal Processing and Two-Stage Localization}

\subsection{AP Received Echo and De-chirp}
We define the target reflection loss from $\boldsymbol{\psi}_{m,n}$ as $\vert\Gamma_{m,n}|^2\in[0,1]$, given by\cite{he2026rss, 11482103}: $|\Gamma_{m,n}|=\epsilon_a\theta_{m,n}+\epsilon_b$, where the incident angle $\theta_{m,n}$ is computed by $\theta_{m,n}=\frac{1}{2}\arccos\Big(\frac{\Vert\boldsymbol{\psi}_{m,n}-\boldsymbol{\psi}^{\text{AP}}\Vert^2-\Vert\boldsymbol{\psi}_{m,n}-\boldsymbol{\psi}^{s}\Vert^2 -\Vert\boldsymbol{\psi}^s-\boldsymbol{\psi}^{\text{AP}}\Vert^2 }{-2\Vert\boldsymbol{\psi}_{m,n}-\boldsymbol{\psi}^{s}\Vert\Vert\boldsymbol{\psi}^s-\boldsymbol{\psi}^{\text{AP}}\Vert}\Big)$. It is worth noting that $\epsilon_a$ and $\epsilon_b$ are the reflection coefficients determined by the permittivity of the reflecting surface. The validity of the reflection loss model has been verified in \cite{9013236}. Furthermore, the target-reflected  echo signal received at the AP is given by
\begin{align}
     y_{s}(t)
&=\sum\nolimits_{m=1}^{M}\sum\nolimits_{n=1}^{N}\xi^s_{m,n} s(t-\tau^s_{m,n}) +n_s(t).
\end{align}
Here,  $\xi^s_{m,n}=\sqrt{P_t}\alpha_{m,n}
[\qh_{s,m}]_n^{*}
[\qg_{m}]_n
w_m\vert\Gamma_{m,n}\vert^2 $ is the sensing effective cascaded gain, while $n_s(t)$ is the AWGN with power $\sigma_s^2$ at time slot $t$. Moreover, $\tau_{m,n}^s$ is the round trip propagation delay associated with the path from the $n$-th PA on the $m$-th waveguide to the target and then to the AP. The sensing SNR at AP is $   \text{SNR}_s=\frac{\vert \sum_{m=1}^{M}\sum_{n=1}^{N}\xi^s_{m,n}\vert^2}{\sigma_s^2}. $

\begin{figure*}[!t]
\begin{equation}\label{eq:linear}
2(x^{\mathrm{AP}}-x^{p}_k)x^{s}
+2(y^{\mathrm{AP}}-y^{p}_k)y^{s}
+2(z^{\mathrm{AP}}-d)z^{s}
+2r_k d_{\text{AP}}
+m_{\mathrm{TSoA},k}
=
r_k^2
+(x^{\mathrm{AP}})^2
+(y^{\mathrm{AP}})^2
+(z^{\mathrm{AP}})^2
-(x^{p}_k)^2
-(y^{p}_k)^2
-d^2.\tag{22}
\end{equation}
\vspace{-0.5em}
\hrule
\vspace{-1.6em}
\end{figure*}

\vspace{-0.5em}
\subsection{Stage I: Delay Estimation}

The received echo is mixed with the local reference to generate the beat frequency signal, which is 
\begin{equation}\label{eq:beat_signal}
    b(t)=y_{s}(t) s^*(t).
\end{equation}
Although the received echo is a superposition of the signals from all activated PAs, the per-PA components can  be distinguished during localization due to their different propagation delays\cite{he2026rss}.
By substituting the expression of  $s(t-\tau^s_{m,n})$ into \eqref{eq:beat_signal}, the beat signal can be expanded as
\vspace{-0.2em}
\begin{equation}\label{eq:beat_signal2}
    b(t)=\sum\nolimits_{m=1}^{M}\sum\nolimits_{n=1}^{N}\xi_{m,n}^s e^{j\phi^c_{i}}e^{-j2\pi f_{b,n}t}e^{j\vartheta_{m,n}^c} + \tilde n(t),
\end{equation}
where the de-chirped noise is $\tilde n(t)=n_s(t)s^*(t)$ and $\vartheta_{m,n}^c \triangleq \pi\!\left(u(\tau_{m,n}^s)^2 - 2f_0\tau_{m,n}^s\right)$ denotes the delay-dependent constant phase term, and $f_{b,m,n} \triangleq u\tau_{m,n}^s$ represents the corresponding beat frequency determined by the propagation delay.  Within one symbol duration, the first exponential term is constant. The modulation phase mainly introduces a constant phase rotation and slight spectral spreading, while the signal energy remains concentrated around $f_b$. Therefore, performing a fast Fourier transform (FFT) and peak detection on the beat signal still enables robust estimation of $f_{b,m,n}$  and $\hat{\tau}^s_{m,n}=\hat{f}_{b,m,n}/u$. Finally, we obtain the sensing link propagation distance $r_{m,n}=\hat{\tau}^s_{m,n}c$.

\begin{remark}
In our considered activation pattern, the delay differences between adjacent activated PAs are well above the minimum resolvable delay. Therefore, echoes from different activated PAs can be reliably resolved as distinct FFT peaks.
\end{remark}

\enlargethispage{-0.03in}

\vspace{-0.3em}
\subsection{Stage II: TSoA Localization}
For notational simplicity, we use a single index $k$ to denote the $k$-th activated PA. The corresponding TSoA observation can then be expressed as
\vspace{-0.2em}
\begin{align}\label{eq:r_mn}
      r_{k}&= \Vert\boldsymbol{\psi}^{s} - \boldsymbol{\psi}^{\text{AP}}\Vert +  \Vert \boldsymbol{\psi}_{k}-\boldsymbol{\psi}^{s}\Vert + n_{\text{TSoA},k} \notag \\
      &=d^t_{k} + d_{\text{AP}} + n_{\text{TSoA},k}, 
\end{align}
where  $d^{t}_{k}$ represents the distance between the $k$-th activated PA and the sensing target, which is given by $d^t_{k}=\sqrt{(x^{\text{p}}_{k}-x^{s})^2+(y^{\text{p}}_{k}-y^{s})^2+(d-z^{s})^2}$, while  $d_{\text{AP}}$ denotes the distance between the sensing target and the AP, given by $d_{\text{AP}}= \sqrt{(x^{s}-x^{\text{AP}})^2+(y^{s}-y^{\text{AP}})^2+(z^{s}-z^{\text{AP}})^2}$.

Accordingly, the least-squares estimator for the target estimation can be formulated as
\vspace{-0.2em}
\begin{equation}
\hat{\boldsymbol{\psi}}^{s}
=
\arg\min_{\boldsymbol{\psi}^{s}}
\sum\nolimits_{k=1}^{K}
\left(r_{k} - d^t_{k}
-d_{\text{AP}}\right)^2.
\end{equation}
It  is observed that the nonlinearity of the TSoA-based localization problem arises from the two square root distance terms. By rearranging the range measurement in \eqref{eq:r_mn}, the PA-target distance is computed by
\vspace{-0.2em}
\begin{equation}
r_{k}-d_{\text{AP}}=d^{t}_{k}+n_{\mathrm{TSoA},k}.
\end{equation}
We square both sides to derive a tractable linearized representation as
\vspace{-0.2em}
\begin{equation}
r_k^2-2r_k d_{\mathrm{AP}}+d_{\mathrm{AP}}^2
=
(d_k^t)^2+n_{\mathrm{TSoA},k}^2+2d_k^t n_{\mathrm{TSoA},k}.
\end{equation}
For compact notation, we define the noise-induced residual as
\vspace{-0.2em}
\begin{equation}
m_{\mathrm{TSoA},k}
=
n_{\mathrm{TSoA},k}^2+2d_k^t n_{\mathrm{TSoA},k}.
\end{equation}
Accordingly, the squared measurement equation is
\vspace{-0.2em}
\begin{equation}
r_k^2-2r_kd_{\mathrm{AP}}+d_{\mathrm{AP}}^2
=
(d_k^t)^2+m_{\mathrm{TSoA},k}.
\end{equation}

Then, by substituting the geometric expressions of $d_{\mathrm{AP}}^2$ and $(d_k^t)^2$, and collecting the terms related to the unknown target position, the linearized measurement model in \eqref{eq:linear}, shown at the top of the page.
Let
$
\boldsymbol{\theta}^{\text{s}}
=
\begin{bmatrix}
x^s , y^s , z^s , d_{\text{AP}}
\end{bmatrix}^{T} \in \mathbb{R}^{4\times1}
$
be the augmented unknown vector. Then, the linearized TSoA equation can be written as $\mathbf L \boldsymbol{\theta}^{\text{s}} + \mathbf q = \mathbf b$, where
\vspace{0em}
\setcounter{equation}{22}
\begin{equation}
\mathbf L
=
2\begin{bmatrix}
x^{\mathrm{AP}}-x^{p}_1 & y^{\mathrm{AP}}-y^{p}_1 & z^{\mathrm{AP}}-d & r_1 \\
\vdots & \vdots & \vdots & \vdots \\
x^{\mathrm{AP}}-x^{p}_{k} & y^{\mathrm{AP}}-y^{p}_{k} & z^{\mathrm{AP}}-d & r_{k} \\
\vdots & \vdots & \vdots & \vdots \\
x^{\mathrm{AP}}-x^{p}_{K} & y^{\mathrm{AP}}-y^{p}_{K} & z^{\mathrm{AP}}-d & r_{K}
\end{bmatrix}\in\mathbb{R}^{K\times4},
\end{equation}
\begin{equation}
\mathbf q
\!=\!
\begin{bmatrix}
m_{\mathrm{TSoA},1}, \cdots, m_{\mathrm{TSoA},k}, \cdots, m_{\mathrm{TSoA},K}
\end{bmatrix}^{T}
\!\!\!\in\! \mathbb{R}^{K\times1},
\end{equation}
\begin{equation}\small
\mathbf b
=
\begin{bmatrix}
r_1^2\!+\!(x^{\mathrm{AP}})^2\!+\!(y^{\mathrm{AP}})^2\!+\!(z^{\mathrm{AP}})^2\!-\!(x_1^{p})^2\!-\!(y_1^{p})^2\!-\!d^2\\\vdots \\
r_k^2\!+\!(x^{\mathrm{AP}})^2\!+\!(y^{\mathrm{AP}})^2\!+\!(z^{\mathrm{AP}})^2\!-\!(x_k^{p})^2\!-\!(y_k^{p})^2\!-\!d^2 \\
\vdots \\
r_{K}^2\!+\!(x^{\mathrm{AP}})^2\!+\!(y^{\mathrm{AP}})^2\!+\!(z^{\mathrm{AP}})^2\!-\!(x_{K}^{p})^2\!-\!(y_{K}^{p})^2\!-\!d^2
\end{bmatrix}\!\!\in\! \mathbb{R}^{K\times1}.
\end{equation}

Since the TSoA observations associated with different activated PAs generally have unequal estimation accuracy, a weighted linear least squares estimator (WLLS)\footnote{{The WLLS estimator in \eqref{WLLS} requires $\operatorname{rank}(\mathbf L)=4$, with positive weights and $K\geq4$. For the considered coplanar PA deployment, this requires $z^{\mathrm{AP}}\neq d$, non-collinear activated PA positions, and a range vector that is not an affine function of the horizontal PA coordinates; otherwise, $\operatorname{rank}(\mathbf L)<4$.}} is adopted as 
\vspace{-0.1em}
\begin{equation}\label{WLLS}
    \hat{\boldsymbol{\theta}}^{\text{s}}=(\qL^T \qW_s \qL)^{-1}\qL^T \qW_s\qb.
\end{equation}
Here,  $\qW_s=\operatorname{diag}(w_{s,1},\dots,w_{s,K})$, where $w_{s,k}$ represents the normalized weight associated with the $k$-th  activated PA, obtained by dividing its sensing SNR by the maximum sensing SNR among all activated PAs.
Finally, the sensing target position is obtained from the first three elements of $\hat{\boldsymbol{\theta}}^{\text{s}}$.

\vspace{-2em}
\subsection{CRB Analysis}
By stacking the observations of all activated PAs into a vector, we have
\begin{equation}
    \qr_{\text{TSoA}} = \qf_{\text{TSoA}}(\bm{\psi}^s) + \qn_{\text{TSoA}}.
\end{equation}
Moreover, the noise is assumed to be a zero-mean Gaussian vector with  covariance matrix $    \qC=\operatorname{diag}(\sigma^2_1,\sigma^2_2,\dots,\sigma^2_K
)$, where the corresponding measurement noise variance is given by \cite{he2026stochastic}:  $\sigma_k^2 = \frac{c^2}{8\pi^2f_0W_b N_s\gamma_{s,k}}$,  where $W_b$ is the effective bandwidth, $N_s$ is the number of transmited symbols and $\gamma_{s,k}$ denotes the sensing SNR corresponding to the $k$-th PA link. In the CRB derivation, $\sigma_k^2$ is assumed to be known and is evaluated from the measurement stage.
Therefore, the probability density function (PDF) of $\qr_{\text{TSoA}}$   is given by
\vspace{-0.3em}
\begin{equation}
    p(\qr_{\text{TSoA}})
   \! =\!
    \frac{1}{(2\pi)^{\frac{K}{2}}\prod_{k=1}^{K}\sigma_k}
    \operatorname{exp}\!\bigg(\!\!\!
    -\!\frac{1}{2}\!
    \sum_{k=1}^{K}
    \frac{\big(r_k-d_k^t-d_{\text{AP}}\big)^2}{\sigma_k^2}
    \!\bigg).
\end{equation}

\noindent The Fisher information matrix (FIM) is then
\vspace{-0.2em}
\begin{equation}
    \qJ(\bm{\psi}^s)
    =
    \bigg(
    \frac{\partial\qf_{\text{TSoA}}}{\partial\bm{\psi}^s}
    \bigg)^T
    \qC^{-1}
    \bigg(
    \frac{\partial\qf_{\text{TSoA}}}{\partial\bm{\psi}^s}
    \bigg),
\end{equation}

\noindent while the entries of $\qJ(\bm{\psi}^s)$ are
\begin{equation}
    \qJ(\bm{\psi}^s)=\begin{bmatrix}
J_{xx} & J_{xy} & J_{xz} \\
J_{yx} & J_{yy} & J_{yz} \\
J_{zx} & J_{zy} & J_{zz}
\end{bmatrix},
\end{equation}
\vspace{-0.1cm}
\begin{align}
    J_{xx}
    &=
    \sum\nolimits_{k=1}^K
    \frac{1}{\sigma^2_k}
    \left(
    \frac{x^s-x^p_k}{d^t_k}
    +
    \frac{x^s-x^{\text{AP}}}{d_{\text{AP}}}
    \right)^2,\\
    J_{yy}
    &=
    \sum\nolimits_{k=1}^K
    \frac{1}{\sigma^2_k}
    \left(
    \frac{y^s-y^p_k}{d^t_k}
    +
    \frac{y^s-y^{\text{AP}}}{d_{\text{AP}}}
    \right)^2,
    \end{align}
    \begin{align}
    J_{zz}
    &=
    \sum\nolimits_{k=1}^K
    \frac{1}{\sigma^2_k}
    \left(
    \frac{z^s-d}{d^t_k}
    +
    \frac{z^s-z^{\text{AP}}}{d_{\text{AP}}}
    \right)^2,\\  
    J_{xy}\!&=\!J_{yx}\!
    =\!
    \sum\nolimits_{k=1}^K
    \frac{1}{\sigma^2_k}
    \left(
    \frac{x^s\!-x^p_k}{d^t_k}
   \! +\!
    \frac{x^s\!-x^{\text{AP}}}{d_{\text{AP}}}
    \right)\!\!\! \notag \\
    &\times\left(
    \frac{y^s\!-y^p_k}{d^t_k}
    \!+\!
    \frac{y^s\!-y^{\text{AP}}}{d_{\text{AP}}}
    \right),
         \end{align}
    \begin{align}
    J_{xz}\!&=\!J_{zx}\!
    =\!
    \sum\nolimits_{k=1}^K
    \frac{1}{\sigma^2_k}
    \left(
    \frac{x^s\!-x^p_k}{d^t_k}
   \! +\!
    \frac{x^s\!-x^{\text{AP}}}{d_{\text{AP}}}
    \right)\!\!\! \notag \\
    &\times\left(
    \frac{z^s\!-d}{d^t_k}
    +
    \frac{z^s\!-z^{\text{AP}}}{d_{\text{AP}}}
    \right),
        \end{align}
    \begin{align}
    J_{yz}\!&=\!J_{zy}\!
    =\!
    \sum\nolimits_{k=1}^K
    \frac{1}{\sigma^2_k}
    \left(
    \frac{y^s\!-y^p_k}{d^t_k}
    \!+
    \frac{y^s\!-y^{\text{AP}}}{d_{\text{AP}}}
    \right) \notag \\
    &\times\left(
    \frac{z^s\!-d}{d^t_k}
   \! +
    \frac{z^s\!-z^{\text{AP}}}{d_{\text{AP}}}
    \right).
\end{align}
The CRB for estimating the target position is
\begin{equation}
    \mathbf{CRB}=\qJ^{-1}(\bm{\psi}^s).
\end{equation}
Thus, we  obtain the corresponding position error bound (PEB): 
\begin{equation}
    \mathrm{PEB}=\sqrt{\mathrm{Tr}(\mathbf{CRB})},
\end{equation}
which represents the theoretical lower bound on the target localization error. 
To evaluate the practical localization performance, the root mean square error (RMSE) is used, and is defined as 
$\mathrm{RMSE}=\sqrt{\mathbb{E}\big\{\big\|\hat{\boldsymbol{\psi}}^{s}
-\boldsymbol{\psi}^{s}\big\|^{2}\big\}}$. 

\begin{remark}
For the TSoA measurement vector in (27) and a given sensing geometry, the measurement noise variance is inversely proportional to the per-link sensing SNR defined in (14). Thus, a higher sensing SNR generally leads to a lower CRB/PEB, motivating its adoption as a tractable sensing metric.
\end{remark}

\vspace{-0.2cm}

\begin{figure*}[t]
\vspace{-1em}
    \centering
    \begin{minipage}[t]{0.3\textwidth}
        \centering
        \includegraphics[trim=0 0cm 0cm 0cm,clip,width=1.1\textwidth]{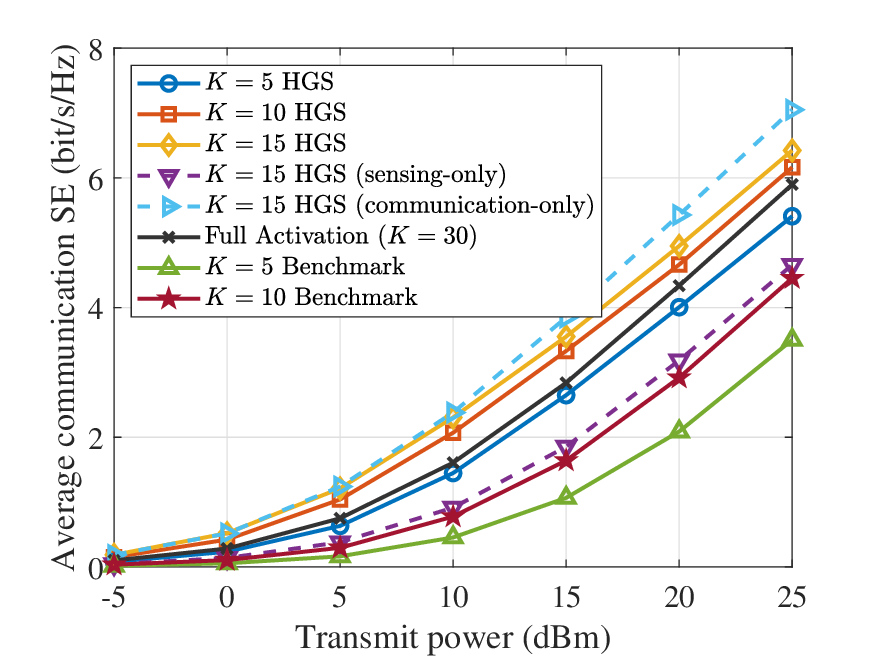}
        \vspace{-1.9em}
        \caption*{\small \textbf{Fig. 2:} Average  SE versus the transmit power. }
        \label{fig:PtSE}
    \end{minipage}
    \hfill
    \begin{minipage}[t]{0.3\textwidth}
        \centering
        \includegraphics[trim=0 0cm 0cm 0cm,clip,width=1.1\textwidth]{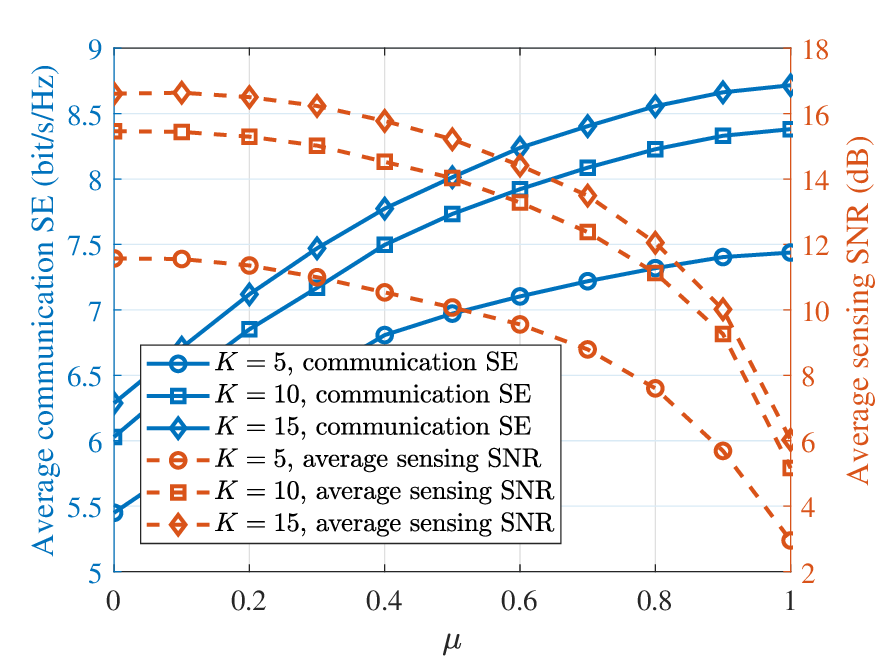}
       \vspace{-1.9em}
        \caption*{\small \textbf{Fig. 3:} Average SE and sensing SNR versus the weighting factor $\mu$. }
        \label{fig:mu}
    \end{minipage}
    \hfill
    \begin{minipage}[t]{0.3\textwidth}
        \centering
        \includegraphics[trim=0 0cm 0cm 0cm,clip,width=1.1\textwidth]{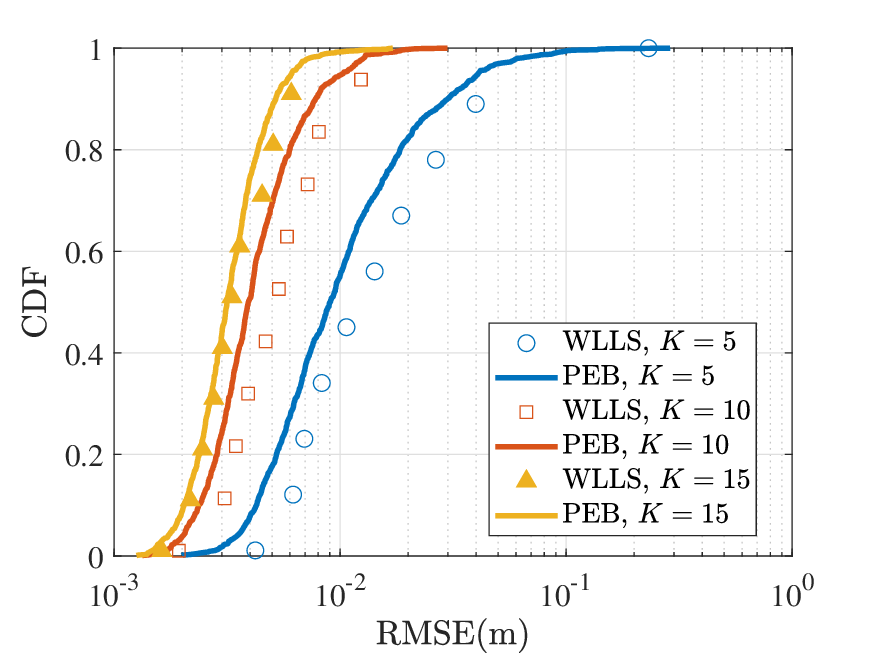}
        \vspace{-1.9em}
        \caption*{\small \textbf{Fig. 4:} CDF of the localization RMSE and PEB for different numbers of activated PAs. }
        \label{fig:CDF}
    \end{minipage}
\vspace{-1.4em}
\end{figure*}

\section{Optimization Problem Formulation}
In this section, we maximize a weighted combination of the communication and sensing metrics through PA activation. Since the activated PA positions significantly affect localization accuracy, a collinear activation pattern provides insufficient geometric diversity, leading to poor localization feasibility~\cite{he2024cramer}. To address this issue, we impose a waveguide activation constraint and introduce the binary variable $\beta_m$ to indicate whether the $m$-th waveguide is active. Following \textit{Remark} 2, the sensing SNR is adopted as a tractable sensing metric~\cite{hu2026pinching}. The activation constraints promote geometric diversity, while WLLS feasibility additionally requires \(\operatorname{rank}(\mathbf L)=4\). The resulting problem is formulated as
\vspace{-0.2em}
\begin{subequations}\label{opt_eq_PA}
\begin{align}
\max_{\alpha_{m,n}, \beta_m}\;
&\mu \text{SE}+(1-\mu)\log_2(\text{SNR}_s) \\
\mathrm{s.t.}\quad
&\sum\nolimits_{m=1}^{M}\sum\nolimits_{n=1}^{N}\alpha_{m,n}=K, \label{PA_const1}\\
&\sum\nolimits_{m=1}^{M}\beta_m \ge 3, \label{PA_const2}\\
&\alpha_{m,n}\le \beta_m,\quad \forall m,n, \label{PA_const3}\\
&\sum\nolimits_{n=1}^{N}\alpha_{m,n}\ge \beta_m,\quad \forall m, \label{PA_const4}\\
&\alpha_{m,n}\in\{0,1\},\quad \forall m,n, \label{PA_const5}\\
&\beta_m\in\{0,1\},\quad \forall m. \label{PA_const6}
\end{align}
\end{subequations}

The objective function
is the weighted SE and sensing SNR. The constraint \eqref{PA_const1} represents  the number of activated PAs. 
Constraints \eqref{PA_const2}-\eqref{PA_const4} promote geometric diversity by requiring PA activation on at least three waveguides, thereby reducing the risk of rank deficiency.
Finally, constraints \eqref{PA_const5} and \eqref{PA_const6} impose binary requirements on the PA activation variables and waveguide activation indicators, respectively.

It is noted that the PA positions are strongly coupled with both the dielectric waveguide propagation and the free-space channel, which makes the objective function  non-convex. Moreover, the binary activation variables introduce a discrete combinatorial structure, which makes the problem intractable for exhaustive search when the system dimension becomes large. Therefore, a heuristic greedy selection (HGS) algorithm is adopted to efficiently determine the PA activation pattern with low computational complexity. Specifically, one PA is first selected from each waveguide to form an initial feasible set, where all cross-waveguide three-PA combinations are examined and the best one is retained. Then, the remaining PAs are added one by one according to the largest objective improvement until $K$ PAs are activated. The resulting computational complexity of the HGS algorithm is approximately 
$\mathcal{O}_{\mathrm{HGS}}\approx\mathcal{O}\left(N^{4}+KN^{2}\right)$, 
which is much lower than  that of the exhaustive search. The steps defined in the  greedy procedure are summarized in \textbf{Algorithm 1}. 
\vspace{-0.3em}
\begin{algorithm}[t]
\footnotesize\caption{HGS algorithm for solving (39)}
\label{alg:greedy_pa}
\begin{algorithmic}[1]
\State \textbf{Input:} $K$, $N$, $M$, $\mathbf{h}_c$, $\mathbf{h}_s$, $\mathbf{G}$, $\mu$, 
\State \textbf{Initialize:} $\boldsymbol{\alpha}^{\star}=\mathbf{0}$, the best objective value $f^{\star}$ as $-\infty$.

\For{$n_1=1$ to $N$}
    \For{$n_2=1$ to $N$}
        \For{$n_3=1$ to $N$}
            \State Construct $\boldsymbol{\alpha}$ by activating
PAs $(1,n_1)$, $(2,n_2)$, and $(3,n_3)$
            \State Evaluate $f(\boldsymbol{\alpha})$
            \If{$f(\boldsymbol{\alpha}) > f^{\star}$}
                \State $\boldsymbol{\alpha}^{\star} \gets \boldsymbol{\alpha}$, \quad $f^{\star} \gets f(\boldsymbol{\alpha})$
            \EndIf
        \EndFor
    \EndFor
\EndFor

\While{$\sum_i \alpha_i^{\star} < K$}
    \State $f_{\mathrm{loc}}=-\infty$,
$\boldsymbol{\alpha}_{\mathrm{loc}}=\boldsymbol{\alpha}^{\star}$
    \For{each inactive PA $k$}
        \State Activate PA $k$ temporarily and obtain $\boldsymbol{\alpha}$
        \State Evaluate $f(\boldsymbol{\alpha})$
\If{$f(\boldsymbol{\alpha})>f_{\mathrm{loc}}$}
    \State $f_{\mathrm{loc}}\gets f(\boldsymbol{\alpha})$,
    $\boldsymbol{\alpha}_{\mathrm{loc}}\gets\boldsymbol{\alpha}$
\EndIf
    \EndFor
\State $\boldsymbol{\alpha}^{\star}
\gets\boldsymbol{\alpha}_{\mathrm{loc}}$
\EndWhile

\State \textbf{Output:} $\boldsymbol{\alpha}^{\star}$
\end{algorithmic}
\end{algorithm}

\vspace{-0.3em}
\section{Numerical Results}
{In this section, the performance of the PASS is evaluated under different network conditions. Specifically, the PASS  consists of \(M=3\) waveguides, with \(N=10\) candidate PAs deployed on each waveguide, resulting in \(30\) candidate PAs in total. The carrier frequency is set to \(28\ \mathrm{GHz}\) and the chirp bandwidth is set to \(B=300\ \mathrm{MHz}\).} The waveguide length, height, and refractive index are set to \(L=40\ \mathrm{m}\), \(d=5\ \mathrm{m}\), and \(n_e=1.4\), respectively, while the waveguides are located at \(y=[5,10,15]\ \mathrm{m}\). The communication and sensing noise powers are set to \(\sigma_c^2=-90\ \mathrm{dBm}\) and \(\sigma_s^2=-100\ \mathrm{dBm}\), respectively. The weighting factor is set to \(\mu=0.5\). The reflection loss parameters are chosen as \(\epsilon_a=0.00572\) and \(\epsilon_b=0.2825\)\cite{he2026rss}. Moreover, all numerical results are averaged over \(1,000\) random realizations. In each realization, the CU is randomly generated within \(x_c\in[5,25]\), \(y_c\in[8,18]\), and \(z_c=1\), while the sensing target is randomly generated within \(x_s\in[10,30]\), \(y_s\in[5,15]\), and \(z_s=2\). The AP location is fixed at \(\boldsymbol{\psi}^{\mathrm{AP}}=[32,8,1]^T\). We consider a random selection scheme as  a benchmark, where $K$ PAs are randomly activated from the candidate PA set while satisfying the same activation constraints.

Figure~2 shows that the average SE increases with the transmit power for all schemes due to the improved received SNR. The communication-only scheme, corresponding to $\mu=1$ in~\eqref{opt_eq_PA}, achieves the highest SE, whereas the sensing-only scheme ($\mu=0$) yields lower SE due to sensing-oriented PA activation. For the proposed HGS scheme, we set $\mu=0.5$. For the same $K$, the proposed HGS consistently outperforms the benchmark, confirming the effectiveness of the PA activation design. The smaller SE gap between $K=10$ and $K=15$ than between $K=5$ and $K=10$ indicates diminishing gains as more PAs are activated. This is mainly due to stronger phase coupling among PA signals, which may limit the coherent combining gain at the CU. The full activation scheme is also outperformed by optimized partial activation with $K=10$, demonstrating that activating all PAs is not always optimal. Thus, the proposed HGS selects PAs with more favorable channel and phase conditions for communication.



Figure~3 illustrates the tradeoff between communication and sensing performance under different values of the weighting factor. As $\mu$ increases, the average  SE gradually improves, while the average sensing SNR correspondingly decreases, which is consistent with the communication-oriented preference of the objective function. In addition, a larger number of activated PAs generally provides better performance for both metrics. The figure also shows that  a moderate $\mu$ enables a desirable tradeoff.

Figure~4 compares the CDF of the WLLS-based localization RMSE and the corresponding PEB under different numbers of activated PAs. It is observed that both the PEB and RMSE curves shift to the left as $K$ increases, indicating that activating more PAs improves the localization accuracy by providing richer geometric diversity. Moreover, the RMSE curves remain above the corresponding PEB curves, which is consistent with the fact that the PEB serves as a theoretical lower bound while the practical estimator suffers from noise and linearization errors. The similar variation trend between RMSE and PEB also validates the effectiveness of the theoretical localization analysis.

\vspace{-0.8em}
\section{Conclusion}
A distributed PASS-assisted ISAC framework was proposed for joint downlink communication and target localization. A unified LFM-based waveform was designed for the PASS architecture. Subsequently, a WLLS estimator and the corresponding CRB were developed for target localization. To balance communication SE and sensing performance, a PA activation problem was formulated and efficiently solved using a low-complexity HGS algorithm. Numerical results revealed that the proposed PA activation scheme outperformed random selection in both communication and localization performance. The results also indicated that activating all PAs was unnecessary, as a selected subset can achieve comparable or even superior performance.

\vspace{-1em}
\bibliographystyle{ieeetr}
\bibliography{Reference}

@STRING{IEEE_J_COM        = "{IEEE} Trans. Commun."}

@STRING{IEEE_GLOBECOM   = "Proc. IEEE GLOBECOM"}

@book{pozar2021microwave,
  title={Microwave Engineering: Theory and Techniques},
  author={Pozar, David M},
  year={2021},
  publisher={John Wiley \& Sons}
}

@ARTICLE{10506595,
  author={Lyu, Zhidong and others},
  journal={J. Lightw. Technol.}, 
  title={Multi-Channel Photonic {THz-ISAC} System Based on Integrated {LFM-QAM} Waveform}, 
  year={Jun. 2024},
  volume={42},
  number={11},
  pages={3981-3988},
  doi={10.1109/JLT.2024.3392282}}

@ARTICLE{10147248,
  author={Xiong, Yifeng and others},
  journal={IEEE Trans. Inf. Theory}, 
  title={On the Fundamental Tradeoff of Integrated Sensing and Communications Under {G}aussian Channels}, 
  year={2023},
  month={Sept.},
  volume={69},
  number={9},
  pages={5723-5751},
  doi={10.1109/TIT.2023.3284449}}

@ARTICLE{11391537,
  author={Qian, Mengyu and Mu, Xidong and You, Li and Matthaiou, Michail},
  journal=IEEE_J_COM, 
  title={Pinching-Antenna-Based Communications: Spectral Efficiency Analysis and Deployment Strategies}, 
  year={2026},
  volume={74},
  number={},
  pages={4758-4771},
  doi={10.1109/TCOMM.2026.3663517}}

@ARTICLE{11371592,
  author={Xu, Xiaoxia and Mu, Xidong and Liu, Yuanwei and Nallanathan, Arumugam},
  journal={IEEE Trans. Wireless Commun.}, 
  title={Joint Transmit and Pinching Beamforming for Pinching Antenna System {(PASS)}: Optimization-Based or Learning-Based?}, 
  year={2026},
  volume={25},
  number={},
  pages={11449-11464},
  doi={10.1109/TWC.2026.3658451}
}

@INPROCEEDINGS{9013236,
  author={Xing, Yunchou and Kanhere, Ojas and Ju, Shihao and Rappaport, Theodore S.},
  booktitle=IEEE_GLOBECOM,
  title={Indoor Wireless Channel Properties at Millimeter Wave and Sub-Terahertz Frequencies},
  year={2019},
  month={Dec.},
  doi={10.1109/GLOBECOM38437.2019.9013236}}

@article{he2024cramer,
  author  = {He, Jiajun and others},
  title   = {{Cramér}--{R}ao lower bound analysis for elliptic localization with random sensor placements},
  journal = {IEEE Trans. Aerosp. Electron. Syst.},
  year    = {2024},
  month   = {Aug.},
  volume  = {60},
  number  = {4},
  pages   = {5587--5595},
  doi     = {10.1109/TAES.2024.3370890}
}

@ARTICLE{he2026rss,
  author={He, Jiajun and others},
  journal={IEEE Trans. Wireless Commun.},
  title={{RSS} Localization in Cell-Free Massive {MIMO}: Algorithms, Analysis, and Implementation},
  year={2026},
  volume={25},
  pages={4021--4037},
  month={Sept.},
  doi={10.1109/TWC.2025.3607721}
}

@ARTICLE{11482103,
  author={He, Jiajun and others},
  journal={IEEE Trans. Wireless Commun.}, 
  title={{RSS}-Based Localization With a Single Receiver: Method and Stochastic Analysis}, 
  year={2026},
  volume={25},
  number={},
  pages={15116-15133},
  doi={10.1109/TWC.2026.3674423}}

@ARTICLE{zhou2024fluid,
  author={Zhou, Liaoshi and others},
  journal={IEEE Wireless Commun. Lett.},
  title={Fluid Antenna-Assisted {ISAC} Systems},
  year={2024},
  month={Dec.},
  volume={13},
  number={12},
  pages={3533--3537},
  doi={10.1109/LWC.2024.3476148}
}

@ARTICLE{11111701,
  author={Wang, Zhaolin and Ouyang, Chongjun and Liu, Yuanwei and Nallanathan, Arumugam},
  journal={IEEE Wireless Commun. Lett.},
  title={Wireless Sensing via Pinching-Antenna Systems},
  year={2025},
  month={Nov.},
  volume={14},
  number={11},
  pages={3475-3479},
  doi={10.1109/LWC.2025.3595160}
  }

@ARTICLE{11372748,
  author={He, Jiajun and Mu, Xidong and Ngo, Hien Quoc and Matthaiou, Michail},
  journal={IEEE Wireless Commun. Lett.},
  title={Pinching-Antenna System-Assisted Localization: A Stochastic Geometry Perspective},
 year={2026},
  volume={15},
  number={},
  pages={1737-1741},
  doi={10.1109/LWC.2026.3661899}}

@INPROCEEDINGS{xue2026pinching,
  author={Xue, Na and He, Jiajun and Matthaiou, Michail},
  booktitle={Proc. IEEE ICC},
  title={Pinching-Antenna-Assisted {ISAC} Based on the Reflection Law},
  year={2026},
  month={May}
}

@ARTICLE{11185315,
  author={Sui, Zeping and Liu, Zilong and Musavian, Leila and Yang, Lie-Liang and Hanzo, Lajos},
  journal={IEEE Trans. Wireless Commun.},
  title={Generalized Spatial Modulation Aided Affine Frequency Division Multiplexing},
  year={2026},
  volume={25},
  number={},
  pages={4658-4673},
  month={Oct.},
  doi={10.1109/TWC.2025.3613062}
}

@ARTICLE{ding2025pinching,
  author    = {Ding, Zhiguo},
  title     = {Pinching-Antenna Assisted {ISAC}: A {CRLB} Perspective},
  journal   = {npj Wireless Technology},
  year      = {2025},
  volume    = {1},
  number    = {1},
  pages     = {4},
  month     = {Oct.},
  publisher = {Nature Publishing Group UK London}
}

@inproceedings{bozanis2025cramer,
  author    = {Bozanis, Dimitrios and Papanikolaou, Vasilis K. and Tegos, Sotiris A. and Karagiannidis, George K.},
  title     = {Cram{\'e}r-{R}ao Bounds for Integrated Sensing and Communications in Pinching-Antenna Systems},
  booktitle = { Proc. IEEE PIMRC},
  month     = {Sept.},
  year      = {2025},
  doi       = {10.1109/PIMRC62392.2025.11274872}
}

@ARTICLE{he2026stochastic,
  author={He, Jiajun and others},
  journal={IEEE Trans. Commun.},
  title={Stochastic Analysis of {Cramér-Rao} Lower Bound for Positioning in {mmWave-THz} {HetNets}},
  year={2026},
  volume={74},
  number={},
  pages={9043--9060}
}

@ARTICLE{lyu2023radar,
  author={Lyu, Zhidong and others},
  journal={IEEE Trans. Microw. Theory Techn.},
  title={Radar-Centric Photonic Terahertz Integrated Sensing and Communication System Based on {LFM-PSK} Waveform},
  year={2023},
  volume={71},
  number={11},
  pages={5019-5027},
  month={Nov.},
  doi={10.1109/TMTT.2023.3267546}
}

@ARTICLE{hu2026pinching,
  author={Hu, Yanglin and Zhang, Tiankui and Xu, Xiaoxia and Liu, Yuanwei},
  journal={IEEE Trans. Veh. Technol.},
  title={Pinching Antenna-Enabled Integrated Sensing and Communication for Low-Altitude {UAV}},
  year={2026},
  volume={75},
  number={6},
  pages={11728-11733},
  month={Jun.},
  doi={10.1109/TVT.2025.3641978}
}

@article{sui2025multifunctional,
  title={Multi-Functional Chirp Signalling for Next-Generation Multi-Carrier Wireless Networks: Communications, Sensing and ISAC Perspectives},
  author={Sui, Zeping and others},
  journal={arXiv preprint arXiv:2508.06022},
  year={2025}
}

\end{document}